\documentclass[twocolumn,aps,prl,groupedaddress]{revtex4}
\usepackage{amssymb}

\usepackage{graphicx}
\usepackage{dcolumn}
\usepackage{bm}
\usepackage{pifont}

\begin{document}

\title{Flow diagram of the metal-insulator transition in two dimensions}
\author{S. Anissimova$^{(a)}$, S.~V. Kravchenko$^{(a)}$, A. Punnoose$^{(b)}$, A.~M. Finkel'stein$^{(c)}$ \& T.~M. Klapwijk$^{(d)}$}
\affiliation{$^{(a)}$Physics Department, Northeastern University, Boston, Massachusetts 02115, USA}
\affiliation{$^{(b)}$Physics Department, City College of the City University of New York, New York, New York 10031, USA}
\affiliation{$^{(c)}$Department of Condensed Matter Physics, Weizmann Institute of Science, Rehovot 76100, Israel}
\affiliation{$^{(d)}$Kavli Institute of Nanoscience, Faculty of Applied Sciences, Delft University of Technology, 2628 CJ Delft, The Netherlands}
\noaffiliation
\noaffiliation
\maketitle

\textbf{The discovery of the metal-insulator transition (MIT) in two-dimensional (2D) electron systems~\cite{kravchenko94} challenged the veracity of one of the most influential conjectures~\cite{abrahams79} in the physics of disordered electrons, which states that ``in two dimensions, there is no true metallic behaviour''; no matter how weak the disorder, electrons would be trapped and unable to conduct a current. However, that theory did not account for interactions between the electrons. Here we investigate the interplay between the electron-electron interactions and disorder near the MIT using simultaneous measurements of electrical resistivity and magnetoconductance. We show that both the resistance and interaction amplitude exhibit a fan-like spread as the MIT is crossed. From these data we construct a resistance-interaction flow diagram of the MIT that clearly reveals a quantum critical point, as predicted by the two-parameter scaling theory \cite{punnoose05}. The metallic side of this diagram is accurately described by the renormalization group theory \cite{punnoose02} without any fitting parameters. In particular, the metallic temperature dependence of the resistance sets in when the interaction amplitude reaches $\gamma_2\approx0.45$ --- a value in remarkable agreement with the one predicted by theory \cite{punnoose02}.}

\begin{figure}
\centering\includegraphics[width=.4\textwidth]{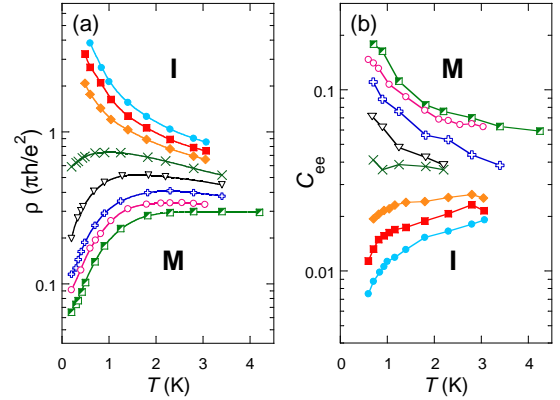}
\caption{Temperature dependences of the resistivity and of the strength of spin-related interactions for electron densities across the MIT. (a): $\protect\rho(T)$ traces at $B=0$.  (b): $C_{ee}(T)$, extracted from the slope of the
magnetoconductance (see text).   Both $\rho(T)$ and $C_{ee}(T)$ display a fan-like structure across the MIT. The densities (in units of $10^{10}$~cm$^{-2}$) are: 7.24, 7.53, 7.83, 8.26, 8.70, 9.14, 9.58, and 9.87.  In panel (a), the densities increase from top to bottom, while in panel (b), they increase from bottom to top, \textit{i.e.}, the interaction increases in the metallic phase and decreases in the insulating phase as the temperature is lowered. \textbf{M} and \textbf{I} regions indicate metallic and insulating phases, respectively.} \label{fig:fan}
\end{figure}

The low amount of disorder in high mobility silicon
metal-oxide-semiconductor field-effect transistors (Si MOSFETs)
allows measurements to be made in the regime of very low electron
densities where correlation effects due to electron-electron
interactions become especially important. (Ratios $r_{s}\equiv
E_{C}/E_{F}>10$ between Coulomb and Fermi energies are easily
reached with Fermi energies of the order $0.7~$meV.) This material
system has the additional advantage that its electron spectrum has
two almost degenerate valleys, which further enhances the
correlation effects. Indeed, the low-temperature drop of the
resistance on the metallic side of the transition
(Fig.\ref{fig:fan}(a)) in Si MOSFETs is the most pronounced among
all 2D electron systems~\cite{kravchenko04,shashkin05}.

At low temperatures $k_BT<\hbar/\tau<E_F$, electrons propagate diffusively. Here, $T$ is temperature, $\tau$ is the electron mean free time, and $E_F$ is the Fermi energy. This region, referred to as the diffusive regime, extends up to a few kelvin near the MIT.  As a result, the low temperature physics of the disordered electron liquid, combined with the large $r_s$ values near the MIT, is determined by the properties of interacting, diffusing electrons~\cite{finkelstein83,castellani84,finkelstein84}.

With this in mind, and with the view of studying the role of
disorder and effective strengths of the electron interactions in
the vicinity of the MIT, we have performed simultaneous
measurements of the resistivity, $\rho$, and in-plane magnetoconductance, $\sigma(B)$.
Since the origin of the magnetoconductance lies in the interaction
corrections to the conductance involving different spin projections
(Hartree-like), the corresponding interaction strength
can be determined from the slope of $\sigma(B)$.
When a magnetic field $B$ is applied parallel to the 2D plane in the
diffusive regime~\cite{tvr82,castellani98}, the magnetoconductance
$\Delta \sigma (B,T)\equiv \sigma (B,T)-\sigma (0,T)$ is
proportional to $b^2$ in the weak field limit $b\equiv g\mu
_{B}B/k_{B}T\ll 1$. On the other hand, at higher temperatures $\hbar
/\tau <k_{B}T<E_{F}$, referred to as the ballistic regime, $\Delta
\sigma (B,T)$ is expected to be proportional to $T b^{2}$
\cite{zala02}. Because the magnetoconductivities in these two
regions are so different, measuring $\Delta \sigma (B,T)$ provides a
reliable way to identify the diffusive region.

\begin{figure*}
\centering \includegraphics[width=.55\textwidth]{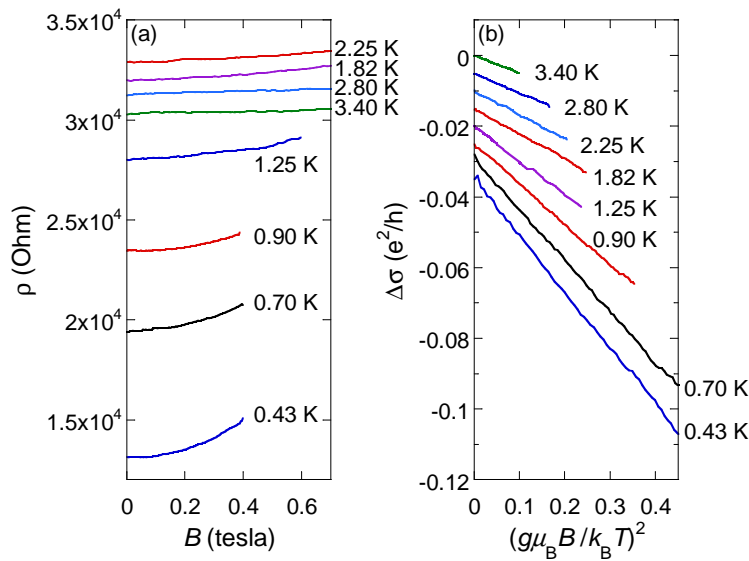}
\caption{Magnetoresistivity and magnetoconductivity for electron density $n_s=9.14\times10^{10}$~cm$^{-2}$ measured at different temperatures.  (a)~Resistivity plotted against magnetic field applied parallel to the 2D plane. (b)~Magnetoconductivity $\protect\Delta\sigma(B,T)\equiv\sigma(B,T)-\sigma(0,T)$ (in units of the quantum conductance) plotted as a function of $(g\protect\mu_B B/k_BT)^2$. The curves in (b) are vertically shifted for clarity. Note that the curves for different temperatures are linear when plotted {\it vs}.\ $(g\protect\mu_B B/k_BT)^2$ and their slopes slowly decrease with temperature.} \label{rawdata}
\end{figure*}

The samples we used had peak electron mobilities
about $3\times10^4$~cm$^2 $/Vs at 0.2~K. The resistivity $\rho$ was
measured by a standard low-frequency lock-in technique. The
magnetoresistance traces, $\rho(B)$, for a representative density in
the diffusive region $n_s=9.14\times 10^{10}$ cm$^{-2}$, are plotted
in Fig.~\ref{rawdata}(a) for different temperatures;
$\rho(T)$ at $B=0$ for this density is the third curve from
the bottom in Fig.~\ref{fig:fan}(a). The corresponding
magnetoconductivities, $\Delta\sigma=1/\rho(B)-1/\rho(0)$, are
plotted as function of $b^{2}$ in Fig.~\ref{rawdata}(b). When
plotted this way, one can see that the $\sigma(b^2)$ curves are
linear. It can also be seen that the slopes decrease slowly with
temperature. This is important, because had the electrons been in
the ballistic regime, the slopes would have instead increased with
temperature by an order of magnitude in the temperature interval
used due to the expected $Tb^2$ dependence. By applying the same
procedure to different electron densities, we find that the
diffusive region extends to approximately $25\%$ in electron
densities above the critical density of the MIT. (The critical
density $n_c\approx 8\times 10^{10}$~cm$^{-2}$ in this sample.)

The explicit form of the slope of $\Delta\sigma(b^2)$ for the single
valley case was derived in Refs.~\cite{tvr82,castellani98}. It is
straightforward to accommodate the valley degrees of freedom. In the
case when there are $n_{v}$ degenerate valleys ($n_{v}=2$ for
silicon), we obtain for $\Delta\sigma(b^2)$, in the limit $b\ll 1$,
the expression
\begin{equation}
\Delta \sigma (b^2)=-0.091\frac{e^{2}}{\pi h}n_{v}^{2}\gamma_{2}\left(\gamma_{2}+1\right)b^{2}.
\label{eqn:linear}
\end{equation}
(We note that the constant $0.091$ in Eq.~(\ref{eqn:linear})
provides a more accurate coefficient compared to the value $0.084$
given in Ref.~\cite{tvr82}.) Here, $\gamma_{2}$ is
the effective electron-electron interaction amplitude in the
spin-density channel. (In standard Fermi-liquid notation,
$\gamma_{2}$ is related to the parameter $F_{0}^{a}$ as
$\gamma_{2}=-F_{0}^{a}/(1+F_{0}^{a})$.) As was observed earlier in
Ref.~\cite{punnoose02}, the large factor $n_{v}^{2}$ due to the
valleys enhances the effect of the $e\text{-}e$ interactions. (In
this paper we have assumed that the valleys are fully degenerate at
all temperatures by restricting ourselves to $T>T_{v}$, where $T_v$
is the temperature scale associated with the valley splitting. This
temperature is not precisely known, but our analysis of the low
temperature data suggests $T_v\approx 0.5$~K. Further details of
this region will be presented elsewhere.)

Eq.~(\ref{eqn:linear}), as was shown in
Ref.~\cite{castellani98}, incorporates renormalization group (RG)
corrections to first order in $\rho$ and is therefore strictly valid
only deep in the metallic region. For general $\rho$, the
magnetoconductance in the diffusive regime will still retain the
$b^2$-form, $\Delta\sigma=-(e^{2}/\pi h)\,C_{ee}(\gamma_2,\rho)\,
n_{v}^{2}\,b^{2}$. The coefficient $C_{ee}$ reflects the strength of
spin-related interactions at \textit{any} value of the resistance
(as long as $g\mu_BB<k_BT<h/\tau$). This is because the in-plane
magnetoconductance is a consequence of the splitting of the spin
subbands. The fluctuations in spin-density lead to finite
temperature corrections to the resistivity \textit{via} the
electron-electron interaction amplitude in the spin-density channel,
$\gamma_2$. Thus, the spin-splitting, by reducing these fluctuations,
leads to a finite magnetoconductance through $\gamma_2$.

The temperature dependences of the parameter $C_{ee}$, extracted
by fitting the $\sigma(b^{2})$ traces in Fig.~\ref{rawdata}(b),
are shown in Fig.~\ref{fig:fan}(b) for various densities across the
MIT. To the best of our knowledge, this is the first observation of
the temperature dependence of the strength of the electron-electron
interactions.
In Fig.~\ref{fig:fan}(a), we plot $\rho(T)$ at zero magnetic field
for the same densities. (In the following, $\rho$ is always
expressed in units of $\pi h/e^2$ \cite{punnoose05}.)
Fig.~\ref{fig:fan} reveals  that not only $\rho$ but also
the interaction strength exhibits a fan-like spread as the MIT is
crossed. We see that, while the interaction grows in the metallic
phase as the temperature is reduced, it is suppressed in the
insulating phase. The magnitudes of these changes depend on how far
the system is from the MIT, with both $\rho$ and $\gamma_2$ becoming
practically temperature independent for densities close to the MIT.
This behaviour is indicative of the flow around a QCP, which is fully
consistent with the theoretical prediction that the evolution of
$\rho$ and the interaction amplitude $\gamma_2$ with temperature are
described by a two-parameter system of coupled RG equations
manifesting an unstable fixed point corresponding to the
QCP~\cite{punnoose05}. Note that a phase diagram can be
presented in different coordinates obtained as a result of various
(rather broad) transformations, preserving the topology of the phase
diagram. In our case, the transformation involves using the slope of
the magnetoconductance.

\begin{figure}
\centering \includegraphics[width=.40\textwidth]{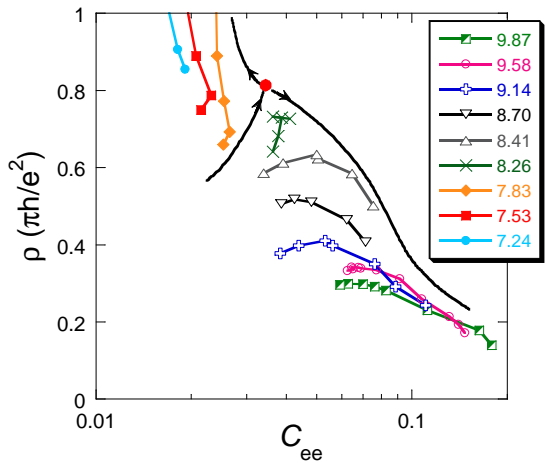}
\caption{The disorder-{interaction} ($\rho$~--~$C_{ee}
$) flow diagram of the 2D electron liquid in Si-MOSFETs. The red dot
at $\rho\approx 0.8$ and $C_{ee}\approx 0.035$ indicates the location of the QCP from which
the three separatrices (black lines) emanate. Arrows indicate the
direction of the flow as the temperature is lowered.  Electron
densities are indicated in units of $10^{10}$~cm$^{-2}$.} \label{fig:flow}
\end{figure}

To see this in a different way, we combine Figs.~\ref{fig:fan}(a)
and (b) and construct a two-parameter ``flow" diagram in the
disorder--interaction ($\rho$~--~$C_{ee}$) space. The flow diagram
is presented in Fig.~\ref{fig:flow}. One can see that such a flow
diagram constructed out of $\rho$ and the parameter $C_{ee}$
(which is sensitive to the amplitude $\gamma_2$) clearly
indicates the existence of a QCP from which three separatrices, one
attractive and two repulsive, can be drawn.

On the metallic side of the flow diagram, a separatrix which forms
the envelope of a family of non-monotonic curves can be clearly seen
in Fig.~\ref{fig:flow} flowing toward the region with $\rho\ll 1$.
The theoretical details of the evolution of $\rho$ and 
$\gamma_2$ in this region with $\rho\ll 1$ were discussed in detail
in Ref.~\cite{punnoose02} in terms of an RG theory.  We recall the
salient features of the theory here. The theory predicts that:\\
%\begin{enumerate}
%\item
(i)~The amplitude $\gamma_2$ increases monotonically as the temperature is reduced.\\
%\item
(ii)~The resistance, on the other hand, has a characteristic non-monotonic form changing from insulating behaviour ($d\rho/dT<0$) at high temperatures to metallic behaviour ($d\rho/dT>0$) at low temperatures, with the maximum value $\rho_{\text{max}}$ occurring at a crossover temperature $T=T_{\text{max}}$.\\
%\item
(iii)~The value of the amplitude $\gamma_2$ at $T=T_{\text{max}}$ is universal in the limit $\rho_{\text{max}}\ll 1$,
depending only on $n_v$; for $n_{v}=1$, it is $2.08$, while for $n_{v}=2$, it has the much lower value $0.45$.\\
%\item
(iv)~Although the values of $\rho_{\text{max}}$ and $T_{\text{max}}$ are not universal and depend on the system, the behaviours of $\rho(T)/\rho_{\text{max}}$ and $\gamma_{2}(T)$ are universal when plotted as functions of $\rho_{\text{max}}$ln$(T/T_{\text{max}})$.
%\end{enumerate}

\begin{figure*}
\centering \includegraphics[width=0.64\textwidth]{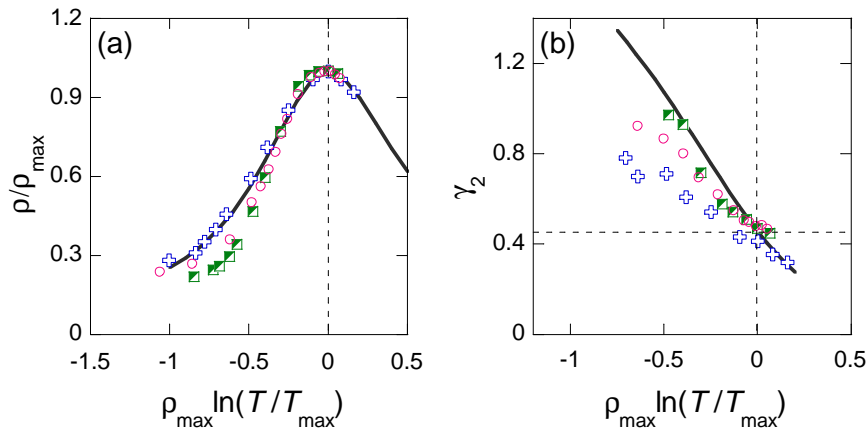}\\
\caption{Comparison between theory (lines) and experiment (symbols).  (a): $\rho/\rho_{\text{max}}$ as a function of $\rho_{\text{max}}\;\text{ln}(T/T_{\text{max}})$.  (b): $\gamma _{2}$ as a function of $\rho_{\text{max}}\;\text{ln}(T/T_{\text{max}})$. Vertical dashed lines correspond to $T=T_{\text{max}}$, the temperature at which $\rho(T)$ reaches  maximum.  Note that at this temperature, the interaction amplitude $\gamma_2\approx 0.45$ (indicated by the horizontal dashed line in (b)), in excellent agreement with theory. Symbols are the same as in Figs.\ref{fig:fan} and \ref{fig:flow} and correspond to $n_{s}=9.87$, $9.58$ and $9.14\times 10^{10}$~cm$^{-2}$.} \label{fig4}
\end{figure*}

The results of comparison between experiment and theory are
presented in Figs.~\ref{fig4}(a) and (b),  for the three largest
densities in the diffusive regime with $\rho_{\text{max}}\alt
0.4~(\pi h/e^2)$. At these densities, the resistance is
small enough, and one can safely extract the interaction amplitudes
$\gamma_2$ from $\Delta\sigma(b^2)$ using Eq.~(\ref{eqn:linear}),
\textit{i.e.}, $C_{ee}=0.091 \gamma_2(1+\gamma_2)$. (As the density
increases, the maximum in $\rho(T)$ becomes less pronounced, and the
non-monotonicity eventually disappears at $n_{s}\agt 1\times
10^{11}$~cm$^{-2}$, as the ballistic region is approached.) The
solid lines are the universal theoretical curves for
$\rho(T)/\rho_{\text{max}}$ and $\gamma_2(T)$ plotted as a function
of $\rho_{\text{max}}\ln(T/T_{\text{max}})$. One sees from
Fig.~\ref{fig4} that at $T=T_{\text{max}}$, the value of the
interaction amplitude $\gamma_2\approx 0.45$, which is in remarkable
agreement with theory for $n_v=2$. The agreement between theory and
experiment is especially striking given that the theory has no
adjustable parameters. Systematic deviations from the universal
curve occur at lower densities as higher order corrections in $\rho$
become important.

The data for $\gamma_{2}$ allow us to calculate the renormalized
Land\'{e} g-factor $g^{\ast }=2(1+\gamma _{2})$.  For our highest
density, it grows from $g^{\ast }\approx2.9$ at the highest
temperature to $g^{\ast }\approx 4$ at the lowest. This is the first
time the g-factor has been measured in the diffusive regime and is
seen to increase with decreasing temperature. Earlier experiments
were done in the ballistic regime where the g-factor was found to be
$g^{\ast}\approx 2.8$ and temperature independent
\cite{anissimova06}.

Finally, we would like to note that the 2D electron system in silicon, used in this study, has a short-range disorder potential and constitutes a multi-valley system: features assumed in the theory.  In other material systems, including GaAs/AlGaAs heterostructures and SiGe heterostructures, the disorder potential is long-range, which leads to very high mobilities.  The diffusive regime in these systems is therefore hard to reach at reasonable temperatures. In addition, high mobilities imply very low carrier densities, which leads to a plethora of finite temperature corrections arising from non-degeneracy effects, large-$r_s$ screening effects, \textit{etc}. Therefore, the transition observed so far in these systems (see, \textit{e.g}., Refs.~\cite{huang06,lai07}) might be a ``high"-temperature crossover phenomenon.  Ultimately, however, at low enough temperatures the diffusive regime will unavoidably be reached. It remains to be seen whether or not the low temperature behaviour in other 2D systems is quantitatively described by the two-parameter scaling theory.

%Correspondence and requests for materials should be addressed to S.V.K.\ at s.kravchenko@neu.edu.

We acknowledge useful discussions with E. Abrahams, C. Castellani, C. Di~Castro and M.~P. Sarachik.  The work at Northeastern University was supported by the NSF grant DMR-0403026.  A.P.\ was supported in part by the PSC-CUNY grant \#60062-3738.  A.F.\ is supported by the Minerva Foundation.

%The experiments were performed by S.A.\ and S.V.K.; A.P.\ and A.M.F.\ were responsible for the theoretical analysis, and T.M.K. has provided the samples.  All authors contributed to writing the manuscript.

\end{document}